# Detecting Disguised Plagiarism

Hatem A. Mahmoud
University of Waterloo
hamahmoud@uwaterloo.ca

**Abstract** Source code plagiarism detection is a problem that has been addressed several times before; and several tools have been developed for that purpose. In this research project we investigated a set of possible disguises that can be mechanically applied to plagiarized source code to defeat plagiarism detection tools. We propose a preprocessor to be used with existing plagiarism detection tools to "normalize" source code before checking it, thus making such disguises ineffective.

## Introduction

Source code plagiarists usually belong to one of two categories: a student plagiarizing programming assignments or a software company plagiarizing the source code of an open source project. Several tools for detecting source code plagiarism already exist, most notable MOSS [7], JPlag [5] and SID [2]. The quality of existing plagiarism detection tools is usually accepted with the assumption that a plagiarist will not have enough time or skills to apply sophisticated code disguises (in the case of a student plagiarist) or that disguising the code effectively may be more expensive than implementing a new program from the scratch (in the case of professional plagiarism). Another assumption is that disguising plagiarized source code will introduce meaningless modifications to the program (e.g., adding useless declarations), which will make the source code look suspicious for any human being investigating it.

Such assumptions do not always hold, mainly for two reasons:

- Existing plagiarism detection tools can be defeated by mechanical disguises; that is, disguises that can be automated. A tool that implements such disguises can be used by unskillful students to hide plagiarism, and can be used by professional plagiarists as a cost-effective means of disguising large plagiarized modules that are comprised of thousands or millions of lines of code.

- Many of those mechanical disguises are still meaningful. They can be applied without making the source code look more suspicious.

In this research project, we investigate such possible disguises, test their effectiveness on existing plagiarism detection tools, propose and develop a preprocessor that deals with most of such disguises and, in order to define the notion of disguised plagiarism precisely, we give definition for a new type of code cloning.

## Plagiarism as Logical Cloning

Firstly the notion of source code plagiarism has to be specified precisely; that is, we need to determine when two pieces of code can be considered disguisedly-plagiarized from one another. Figure 1 shows two sample pieces of code. They represent two implementations in C for a program that computes the permutation of two integers. The two functions have exactly the same logic but were implemented using different language constructs (e.g. while loop instead of for loop, addition to 1 instead of increment, multiplication instead of multiplication assignment, declaration followed by an assignment statement instead of declaration with an initialization … etc). Independent statements and operands were re-ordered, and the trivial disguise of renaming variables was also applied. The question is: can we consider those two pieces of code plagiarized from one another?

```c
long CalcPermutation(long n, long r)
{
      long fact_n = 1, fact_n_r = 1;
      long i, j;

      for(i=2;i<=n;i++)
            fact_n *= i;

      for(j=2;j<=(n-r);j++)
            fact_n_r *= j;

      return fact_n/fact_n_r;
}
```

```c
signed long int ComputePerm(signed long int x,signed long int y)
{
        signed long int counter1;
        signed long int counter2;

        signed long int num;
        signed long int den;

        den = 1;
        counter2 = 2;
        while ( (x-y) >= counter2)
        {
                den = den * counter2;
                counter2 = counter2 + 1;
        }

        num = 1;
        counter1 = 2;
        while (x >= counter1 )
        {
                num = num * counter1;
                counter1 = counter1 + 1;
        }

        return num/den;
}
```

**Figure 1**

For small programs, like simple programming assignments, we usually have a few ways to implement the task. For example, if the assignment to be considered is just about computing permutations, then probably we will get many students having similar or identical logic to that presented in figure 1. Consequently, for small programs, it makes sense not to consider two programs plagiarized from one another unless they share significant textually-equivalent bulks of code, where "textually-equivalent" means identical except for some trivial differences like differences in comments, white spaces or variables' names i.e., changes that do not affect language constructs.

However, for large projects, restricting the definition of plagiarism to textually-equivalent pieces of code is not effective enough. Having two programs, each comprised of hundreds of lines codes, and both having the same logic (though may be implemented using different language constructs and different orderings for independent statements/operands) is something really suspicious; especially since all such differences can be introduced automatically.

We thus propose a new type of code cloning to capture this notion of similarity. We call it *Logical Cloning*. Two pieces of code are said to be logical clones of one another if they have the same logic, even if such logic is implemented using different language constructs and different orderings for statements/operands.

This is to be distinguished from Semantic Cloning (also known as Type IV cloning), which applies to any two pieces of code doing the same functionality i.e., having the same pre and post conditions [6]. Logical cloning, on the other hand, means that the two pieces of code do the same functionality *with the same logic*; they may differ only in the choice of language constructs and the ordering of independent statements and operands.

For example, figure 2 presents a semantic clone for the permutation functions in figure 1. Although the function in figure 2 is quite similar (in appearance) to the function on the left of figure 1, they actually have some differences in logic. The function in figure 2 introduces some improvements that decrease running time.

```
unsigned Permutation(unsigned n, unsigned r)
{
      unsigned fact_n, fact_n_r = 1;
      unsigned i;

      for( i=2 ; i<=(n-r) ; i++)
            fact_n_r *= i;

      for( fact_n=fact_n_r ; i<=n ; i++)
            fact_n *= i;

      return fact_n/fact_n_r;
}
```

**Figure 2**

We consider two programs to be disguisedly-plagiarized from one another if they share significantly large portions of the same logic (logical clones). Note that, according to such definition, two programs do not even have to be implemented in the same programming language in order to be logical clones! Several tools already exist for automatically transforming source code from one programming language into another (typically within the same programming paradigm). For example, Microsoft's Java Language Conversion Assistant converts Java programs to C# automatically.

## Possible Disguises

For the purpose of this research project, we assume that the source code to be checked for plagiarism is written in C. However, most of the disguises mentioned here apply to other languages too. Table 1 presents prominent examples of possible disguises. As mentioned before, acceptable disguises should be mechanical and meaningful; otherwise the number of possible disguises would be infinite.

| Examples of Mechanical Disguises ||
|---|---|
| **Original Code** | **Disguised Code** |
| **1. Re-phrasing control structures** ||
| ```
for(i=0;i<n;i++)
{
   …
}
``` | ```
i=0;
while(i<n)
{
   …
   i++;
}
``` |
| ```
if( x == 'a' )
{ … }
else if ( x == 'b' )
{ … }
else
{ … }
``` | ```
switch( x )
{
case 'a':
   …
   break;
case 'b':
   …
   break;
default:
   …
}
``` |
| **2. Swapping if/else bodies** ||
| ```
if ( condition )
   yes-statement
else
   no-statement
``` | ```
if ( ! condition )
   no-statement
else
   yes-statement
``` |
| **3. Re-phrasing expressions** ||
| `X < Y` | `! ( X >= Y )` |
| `++X` | `X = X + 1` |
| `X += Y` | `X = X + Y` |
| **4. Re-ordering operands** ||
| `X – Y + Z` | `Z + X – Y` |
| **5. Re-distributing operands** ||
| `X * ( Y + Z )` | `X * Y + X * Z` |
| `X && ( Y || Z )` | `X && Y || X && Z` |
| **6. Splitting/Merging statements** ||
| `x+= y = a + b + c, z = n = foo();` | ```
y = a + b + c;
x += y;
n = foo();
z = n;
``` |
| ```
x = getSomeValue();
y = x – z;
``` | `y = (x = getSomeValue()) – z;` |
| `x = (a + b) * c;` | ```
x = a;
x += b;
x *= c;
``` |
| **7. Re-ordering independent statements** ||
| ```
x +=  PI;
y = atan ( 0.9 );
z = x – y;
``` | ```
y = atan ( 0.9 );
x +=  PI;
z = x – y;
``` |

**Table 1**

We tested such disguises by applying them on the source code of a course project that was developed during a computer science graduate course. The project consisted of a few hundreds of lines of code thus provided a reasonable sample. MOSS recognized only 13% similarity between the original code and the disguised one. JPlag gave 10% similarity for the same test case while SID gave 0%.

## Normalization Preprocessor

The reason why such disguises exist at the first place is because of the flexibility of the programming language used, which allows the same logic to be expressed using different language constructs. Such issue can be dealt with by transforming each language construct into a normalized form (canonical form). By transforming source code from its original language (in our case, the C language) into a less-flexible, normalized language, we can eliminate so many disguises that rely on the existence of several equivalent language constructs in the language. The target language is just a subset of the original one (it contains less language constructs) but should be equivalent in power (for each language construct in the original language there exists an equivalent language construct in the target language).

Such transformation will be applied to the source code before examining it for plagiarism i.e., as a preprocessing phase. Existing plagiarism detection tools already have their own preprocessors but they only perform simple types of preprocessing like removing comments and white spaces.

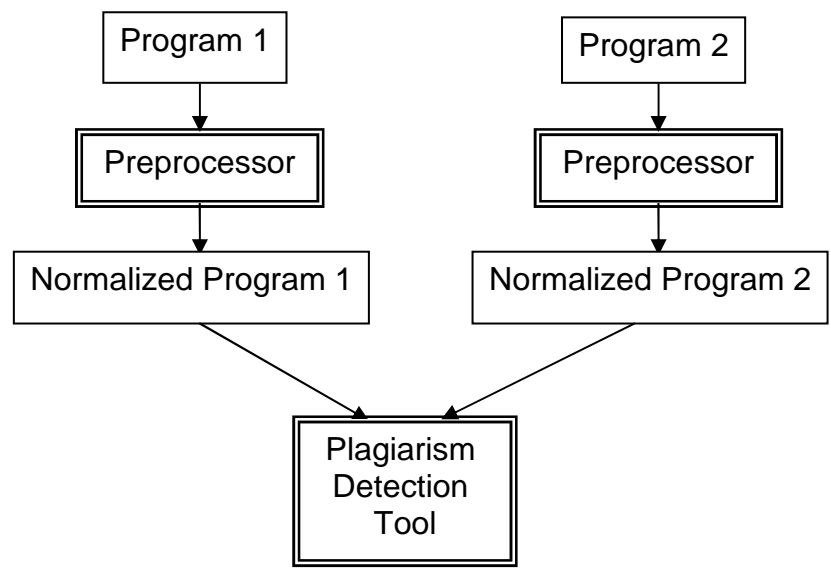

**Figure 3**

We developed a research prototype for such preprocessor for normalizing source code written in C. It handles most of the disguises mentioned in Table 1. However, up till the time of writing this report, some features of the C language are still not supported by our preprocessor.

Table 2 presents some examples of normalization rules for the C language.

| Examples of Normalization Rules | |
|---|---|
| **Original Code** | **Normalized Code** |
| **1. Only one type of loops: while(true) with a break** | |
| `for ( exp1 ; exp2 ; exp3 )`<br>   *body-of-loop* | `exp1;`<br>`while ( true ) {`<br>   `if ( ! exp2 ) { break; }`<br>   *body-of-loop*<br>   `exp3;`<br>`}` |
| `while ( exp )`<br>   *body-of-loop* | `while ( true ) {`<br>  `if ( ! exp ) { break; }`<br>  `body-of-loop`<br>`}` |
| `do`<br>   *body-of-loop*<br>`while( exp );` | `while ( true ) {`<br>  `body-of-loop`<br>  `if ( ! exp ) { break; }`<br>`}` |
| **2. Only two logical operators: ! and &&** | |
| `X || Y` | `! ( ! X && ! Y )` |
| **3. Only two bitwise operators: ~ and &** | |
| `X | Y` | `~( ~X & ~Y )` |
| `X ^ Y` | `~( ~(~X&Y) & ~(X&~Y) )` |
| **4. Only two relational operators: < and ==** | |
| `X > Y` | `Y < X` |
| `X <= Y` | `X < Y  ||  X == Y` |
| `X >= Y` | `Y < X  ||  X == Y` |
| `X != Y` | `! ( X == Y )` |
| **5. No multiple assignments in the same statement** | |
| `x += y = a + b +c, z = n = foo();` | `y = a + b + c;`<br>`x += y;`<br>`n = foo();`<br>`z = n;` |
| **6. Merge successive assignments with the same lvalue** | |
| `x = a;`<br>`x += b;`<br>`x *= c;` | `x = (a + b) * c;` |
| **7. The condition of an if-statement cannot be a negated expression** | |
| `if ( ! condition )`<br>   `block_1`<br>`else`<br>   `block_2` | `if ( condition )`<br>   `block_2`<br>`else`<br>   `block_1` |
| **8. Only one assignment operator: =** | |
| `X += Y` | `X = X + Y` |
| `X >>= Y` | `X = X >> Y` |
| **9. Distribute multiplication over addition whenever possible** | |
| `X * (Y - Z) / K` | `X * Y / K – X * Z / K` |

**Table 2**

Figure 4 shows a block diagram for such preprocessing process. Our prototype performs two passes through the program, one through the source code to parse it into in-memory data structures and another pass through such data structures to generate normalized code. Doing the whole job in a single pass is more efficient, but on the other hand, a two-pass preprocessor may be easier to port to other input languages.

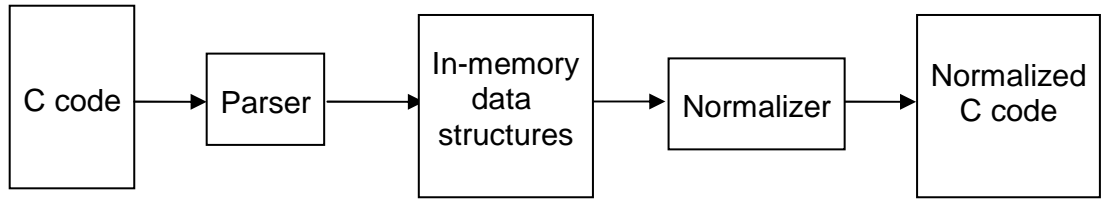

**Figure 4**

The preprocessor was implemented in C++. Table 3 demonstrates the time taken by each of the two passes of the preprocessor as the number of lines of code in the input file increases, though we cannot expect one code file to grow that large. Tests were performed on a machine with 1.7GHz Core Duo CPU and 1GB of RAM. All time measurements are rounded to one decimal place.

| Lines of Code | Pass 1 (Sec) | Pass 2 (Sec) | Total (Sec) |
|---|---|---|---|
| 1000 | 0.1 | 0 | **0.1** |
| 10000 | 0.3 | 0.1 | **0.4** |
| 50000 | 1.4 | 0.7 | **2.1** |
| 100000 | 2.5 | 1.3 | **3.8** |
| 150000 | 4.0 | 2.2 | **6.2** |
| 200000 | 5.1 | 2.5 | **7.6** |
| 250000 | 6.5 | 3.5 | **10** |

**Table 3**

## Discussion

There are some issues to be considered with the proposed approach.

Firstly, the mentioned normalization rules do not deal with re-orderings of independent statements and operands. For the statements re-ordering issue, one possible normalization is to sort mutually independent statements based on some criteria (e.g., type of the statements, number of operators … etc). We have to select our sorting criteria in a manner that minimizes ties as much as possible. However, we have not attempted such idea yet.

Sorting operands based on some criteria may be also possible although the issue of re-ordering operands is not as critical as re-ordering statements. Having resolved other disguises, different orderings of operands will only result in a few token mismatches in a token-based plagiarism test.

GPlag [4] approached the statements re-ordering problem by comparing dependency graphs of individual functions, thus re-formulating the problem of plagiarism detection as a sub-graph isomorphism problem. Unfortunately, the tool is not publicly available for testing.

Another issue with our approach is complexity. It is clear that the proposed approach is too complex to implement. It requires the same effort as developing a language translator that maps one high-level language to another. Consequently, porting such preprocessor to support another language, even having the same paradigm, is not straightforward. Adding, removing or modifying one of the normalization rules requires too much code modifications.

A better approach would be to implement a generic, language-independent preprocessor that obtains its normalization rules from a configuration file. Typically a configuration file will consist of a set of production rules that map language constructs of the source language into language constructs of the target language, although we are not sure yet whether such approach is feasible or not.

## Conclusion

Existing plagiarism detection tools can be defeated using disguises that are both mechanical and meaningful. The main reason why such disguises exist is because of the flexibility of the programming languages used. Normalizing source code into a canonical form can be used to eliminate, or at least minimize, the effectiveness of such disguises. The feasibility of that solution has been proved by developing a preprocessor that performs such normalization. Several enhancements can still be considered to make such approach more useful and applicable.

## References

[1] Brenda S. Baker. *"On finding duplication and near-duplication in large software systems*." In Proceedings of the Second Working Conference on Reverse Engineering. 1995.
[2] Xin Chen and Brent Francia and Ming Li and Brian Mckinnon and Amit Seker. "*Shared Information and Program Plagiarism Detection*." IEEE Transactions of Information Theory. Volume 50. 2004.
[3] Vic Ciesielski, Nelson Wu, and Seyed Tahaghoghi. "*Evolving similarity functions for code plagiarism detection*." In Proceedings of the 10th Annual Conference on Genetic and Evolutionary Computation. 2008.
[4] Chao Liu, Chen Chen, Jiawei Han, Philip S. Yu. " *GPLAG: Detection of Software Plagiarism by Program Dependence Graph Analysis*." In the Proceedings of the 12th ACM SIGKDD International Conference on Knowledge Discovery and Data Mining (KDD'06). 2006.
[5] Lutz Prechelt, Guido Malpohl and Michael Philippsen. "*JPlag: Finding Plagiarisms among a Set of Programs.*" Technical Report 2000-1, Fakultät für Informatik, Universität Karlsruhe.
[6] Chanchal K. Roy, James R. Cordy. "*A Survey on Software Clone Detection Research*." Technical Report. 2007.
[7] Saul Schleimer, Daniel S. Wilkerson, and Alex Aiken. *"Winnowing: local algorithms for document fingerprinting*." In Proceedings of the 2003 ACM SIGMOD international Conference on Management of Data. 2003.